\documentclass[aps,preprint,showpacs,groupedaddress]{revtex4-1}
\usepackage{amssymb}
\usepackage{mathrsfs}
\usepackage{amsmath}
\usepackage{mathtools}
\usepackage{pstricks}
\usepackage{color}
\usepackage{graphicx}
\usepackage{array}
\usepackage{tabularx}
\usepackage{physics}
\usepackage{slashed}
\usepackage{float}

\begin{document}


\title{\bf Nonextensive percolation and Lee-Yang edge singularity from nonextensive $\lambda\phi^{3}$ scalar field theory}




\author{P. R. S. Carvalho}
\email{prscarvalho@ufpi.edu.br}
\affiliation{\it Departamento de F\'\i sica, Universidade Federal do Piau\'\i, 64049-550, Teresina, PI, Brazil}




\begin{abstract}
We compute the critical exponents for nonextensive $\lambda\phi^{3}$ scalar field theory for all loop orders and $|q - 1| < 1$. We apply the results for both nonextensive percolation and Lee-Yang edge singularity problems. The corresponding systems are nonextensive generalizations of their extensive counterparts. For that we employ tools from the recently introduced nonextensive statistical field theory. The results for the nonextensive critical exponents computed depend on the nonextensive parameter $q$, which encodes global correlations among the degrees of freedom of the system. The extensive results are recovered in the limit $q\rightarrow 1$. There is an interplay between global correlations and fluctuations, once the nonextensive critical exponents depend on $q$. This dependence is in agreement with the universality hypothesis.
\end{abstract}

\pacs{47.27.ef; 64.60.Fr; 05.20.-y}

\maketitle


\section{Introduction}\label{Introduction}

\par The higher-order critical behavior of $\phi^{3}$ theory was studied in recent years \cite{KOMPANIETS2021136331,PhysRevD.103.116024}. This model can predict results for the critical behavior of real systems undergoing phase transitions. One example is the ($N + 1$)-state Potts model \cite{potts_1952}. Its one state version, \emph{i. e.} $N = 0$, leads the problem of bond percolation. The two-state model takes into account to the nature of Yang-Lee edge singularities \cite{PhysRev.87.404,PhysRevLett.40.1610}, namely when $N \rightarrow 1$. Now we have to extend this model to the nonextensive realm.

\par The proposal of generalization of both thermodynamics and statistical mechanics for the nonextensive realm  \cite{Tsallis1988} has attracted great interest along the years \cite{Tsallis,TSALLIS1999,TSALLIS2021}. That proposal takes into account global correlations among the degrees of freedom of the system. Such correlations can not be described by Boltzmann-Gibbs statistical mechanics and are encoded by the nonextensive parameter $q$. When the generalization process is employed, the nonextensive parameter $q \in \mathbb{R}$ naturally emerges. It characterizes the nonextensive theory, where for $q < 1$ we have a superextensive theory while for $q > 1$ the theory is subextensive. Some years later the generalization was proposed, it was shown that there is no such an generalization for thermodynamics \cite{PhysRevLett.88.020601}. So nonextensive properties can arise only in the statistical mechanics scenario. As thermodynamic effects are a result of physical phenomena at large scales, the Landau values (which are a result of physical phenomena at that referred scales) for the nonextensive critical exponents must be extensive. Then the nonextensivity of the nonextensive critical exponents will come from their loop corrections, since these corrections are a result of phenomena at small scales. In fact, this behavior was observed in a recent work \cite{submitted}. In that work, both static and dynamic nonextensive critical exponents for $\phi^{4}$ theories were computed from the recently field-theoretic tool for evaluating critical exponents for systems globally correlated, namely nonextensive statistical field theory (NSFT). These critical exponents are responsible for describing the scaling properties of systems with global correlations among their degrees of freedom near a continuous phase transition. The numerical values for these exponents for the particular case of the two-dimensional Ising model were compared with their counterparts obtained from computer simulations and the agreement was satisfactory. One of the great difficulties of motivating the introduction of nonextensive theories is to give the physical interpretation of $q$. In the case just mentioned, this physical interpretation was given in both a direct and natural way. The critical exponents depend on universal properties of the system as its dimension $d$, $N$ and symmetry of some $N$-component order parameter and furthermore if the interactions among their degrees of freedom are of short- or long-range type. They do not depend on nonuniversal properties of the system as its critical temperature, form of its lattice, the nature of spacetime where the field is embedded, as if it is curved \cite{Costa_2019} or a Lorentz-violating one \cite{Carvalho2017} for example. The nonextensive parameter was physically interpreted as a generalized type of interaction, since they represent the global correlations among the many degrees of freedom of the system. Then the critical exponents must depend on $q$. This is in agreement with the universality hypothesis. Also it was found another property of these critical exponents. The effect of the global correlations are slight. In fact, their dependence on $q$ is slight \cite{submitted}. This property could be inferred by considering the following expansion: $[\exp_{2 - q}(-a)]^{2 - q} = e^{-a}\left[1 - \frac{a(a - 2)z}{2} + \frac{a^{2}(3a^{2} - 20a + 24)z^{2}}{24} - \cdot\cdot\cdot \right]$, where $z = q - 1$ and $a = \mathcal{H}/k_{B}T$, for $|q - 1| < 1$, where $\exp_{q}$ is introduced in \cite{Tsallis1988}. We can choice the term $q - 1$ as the expansion parameter, since it is a small on in the range $0.5 < q < 1$. We observe that the first term gives the known Boltzmann-Gibbs results \cite{Gibbs}. The slight corrections now are expressed through loop corrections. They are a result of the fluctuating properties of the system at short distances. Then we make an analogy with that systems encompassed by $\phi^{4}$ model and we expect a similar behavior between the former and the latter ones approached in this work. We hope that the slight corrections will appear as loop corrections. The values of $q$ considered were that contained in the range $0.5 < q < 1$, once continuous phase transitions were found along the ferro-paramagnetic frontier was discovered experimentally  from computer simulations at this interval. The search for such an transition in that range of $q$ was motivated for the discovery of the experimental nonextensive behavior of some maganites, namely La$_{0.7}$Sr$_{0.3}$MnO$_{3}$ and La$_{0.60}$Y$_{0.07}$Ca$_{0.33}$MnO$_{3}$ \cite{PhysRevB.73.092401,PhysRevB.68.014404,Reis_2002,PhysRevB.66.134417}. There has been also many studies of nonextensive effects in the high energy regime. Experimental data obtained from quantum chromodynamics (QCD) from hadron spectra at very high-energies \cite{PhysRevLett.105.022002,PhysRevD.91.114027,PhysRevD.87.114007}, the fractal structure of QCD \cite{PhysRevD.101.034019,physics2030026,PhysRevD.93.054001} and magnetoconductance fluctuations in mesoscopic systems \cite{PhysRevE.104.054129} were observed to be nonextensive. The agreement between experimental data and theoretical predictions for low-energy systems was also found for anomalous diffusion in granular media \cite{PhysRevLett.115.238301} and the transport of cold atoms in dissipative optical lattices \cite{Lutz2013}. As there are so many systems which presented nonextensive behavior, we expect to find a similar behavior for the $\lambda\phi^{3}$ model as well. For studying such behaviors, we employ NSFT tools in the MS scheme \cite{'tHooft1972189}.

\section{Nonextensive critical exponents}

\par For computing the nonextensive critical exponents for the models approached in this work, we have to apply NSFT tools \cite{submitted} for the corresponding Euclidean interacting generating functional for each theory in the path integral formalism. The $\phi^{3}$ model is described by the Eulidean (suitable for statistical systems computation) Lagrangian density
\begin{eqnarray}\label{bare Lagrangian density}
\mathcal{L} =  \frac{1}{2}\partial_{\mu}\phi\partial^{\mu}\phi + \frac{1}{2}m^{2}\phi^{2} + \frac{\lambda}{3!}\phi^{3},
\end{eqnarray}
where $\phi$, $m$ and $\lambda$ are the field, mass and coupling constant of the theory \cite{Amit_1976}. The upper critical dimension of this theory is $d_{c} = 6$, so the natural expansion parameter is $\epsilon = 6 - d$. The corresponding generating functional for the interacting Euclidean field theory, for $q < 1$, is given by \cite{submitted}
\begin{eqnarray}\label{huyhtrjisd}
Z[J] = \mathcal{N}^{-1}\left\{\exp_{2 - q}\left[-\int d^{d}x\mathcal{L}_{int}\left(\frac{\delta}{\delta J(x)}\right)\right]\right\}^{q} \int\exp\left[\frac{1}{2}\int d^{d}xd^{d}x^{\prime}J(x)G_{0}(x-x^{\prime})J(x^{\prime})\right], \nonumber \\
\end{eqnarray}
where $\exp_{q}(-x) = [1 - (1 - q)x]^{1/(1 - q)}$ is the $q$-exponential function \cite{Tsallis1988}, rather than the conventional one $e^{-x}$ \cite{ZinnJustin,Amit}, for which $q \rightarrow 1$. The aforementioned $q$-distribution is a normalizable and thus a physical one for the desired range $q < 1$. We have applied the consistent form of the $q$-distribution \cite{TSALLIS1998534}, namely Eq. (\ref{huyhtrjisd}), one which is raised to a power of the nonextensive $q$-parameter. In Eq. (\ref{huyhtrjisd}),  the constant $\mathcal{N}$ is obtained through the condition $Z[J=0] = 1$, the free propagator is equal to $G_{0}^{-1}(k) = k^{2} + m^{2}$ and is the same as that of the extensive theory \cite{submitted} and $J(x)$ is some auxiliary external source to generate the correlation functions and to be vanished at the end of calculation. By applying the MS scheme \cite{'tHooft1972189}, we obtain, independently, the following all-loop order critical exponents
\begin{eqnarray}\label{eta}
\eta_{q} = \eta - (1 - q)\frac{\alpha(\alpha - 8\beta)}{3(\alpha - 4\beta)[\alpha - 4\beta(2 - q)]}\epsilon, 
\end{eqnarray}
\begin{eqnarray}\label{nu}
\nu_{q}^{-1} = \nu^{-1} - (1 - q)\frac{5\alpha(\alpha - 8\beta)}{3(\alpha - 4\beta)[\alpha - 4\beta(2 - q)]}\epsilon,
\end{eqnarray}
\begin{eqnarray}\label{omega}
\omega_{q} = \omega .
\end{eqnarray}
The aforementioned critical exponents values are valid for any loop level, where $\eta$ and $\nu$ are the all-loop extensive values for these exponents. In fact, the theory is renormalizable for general values of $q$ up to one-loop order. So at this order, the critical exponents depend on $q$. Therefore when we proceed to renormalize the theory for two-loop order and beyond, it is not renormalizable for any value of $q$ but only for the extensive value $q = 1$. Then, for two-loop level and higher ones, the critical exponents values do not depend on $q$ and assume the same values as their extensive counterparts. Thus, the nonextensive critical exponents are composed of $q$-independent tree-level, two-loop and higher order contributions and some one-loop level $q$-dependent term. In fact, this behavior in which a given field theory is renormalizable at leading order one-loop order for any values of some characteristic parameter ($q$ in our case) and nonrenormalizable for two-loop level higher ones general values of that parameter, but is renormalizable just for some fixed value of the parameter for two-loop level higher ones is not a new feature in literature. That is the case of quantum field theories in curved spacetime for example \cite{PhysRevD.14.1965,PhysRevLett.54.2281}. The remaining nonextensive critical exponents values can be obtained through scaling relations among them \cite{ZinnJustin,Amit}. We observe that the Landau values for the nonextensive critical exponents just computed are extensive, \emph{i. e.}, they do not depend on $q$. As they can be also evaluated from thermodynamic theory, by considering just large length scales, their independence on $q$ confirms the general result of Ref. \cite{PhysRevLett.88.020601} as discussed in Sec. \ref{Introduction}. Furthermore, the $q$-dependence of the nonextensive critical exponents is slight, as also discussed in Sec. \ref{Introduction}, since it comes from a one-loop term. These exponents satisfy to the universality hypothesis, once the $q$ parameter encodes global correlations or equally represents effective interactions among the degrees of freedom of the system. Then, the universality hypothesis implies that critical indices must depend on $q$. Another way of interpreting this result is that the nonextensive indices have to depend on $q$, because the nonextensive statistical weight of Eq. (\ref{huyhtrjisd}) produces a modification of the internal properties
of the field such that it modifies the way it interacts (with some dependence on $q$) with itself. From this fact, emerges some relation between fluctuations and global correlations. This mechanism is one that occurs in the internal space of the field and not in the spacetime where it is embedded. Then, this leads, necessarily, to $q$-dependent critical indices. Now we have to present the numerical results for the nonextensive critical exponents for percolation and Lee-Yang edge singularity problems.

\subsection{Percolation}

\par For the percolation problem, $\alpha = -1$ and $\beta = -2$ \cite{Bonfirm_1981} in Eqs. (\ref{eta})-(\ref{nu}). Then we compute the independent critical indices $\eta_{q}$, $\nu_{q}$ and $\omega_{q}$ and the remaining ones through the scaling relations among them \cite{PhysRevD.103.116024}. In table \ref{exponentsP2} we present the results for the numerical values for the nonextensive critical exponents for the percolation problem for $2d$ systems. For that, we apply the corresponding extensive exponents values obtained from the exact solution \cite{PhysRevD.103.116024}. In this case, these results for $2d$ systems are \emph{exact}.  

\begin{table}[H]
\caption{Exact nonextensive critical exponents, for some values of $q$, to $2$d nonextensive systems in percolation, obtained from NSFT.}
\begin{tabular}{ p{1.4cm}p{2.2cm}p{2.2cm}p{2.2cm}  }
 \hline
 \hline
 $q$ & $\alpha_{q}$ & $\beta_{q}$ & $\gamma_{q}$   \\
 \hline
 1\cite{PhysRevD.103.116024}   &  -0.666  &  0.139 &  2.389          \\
 0.9  &  -0.144  &  0.131 &  1.881    \\
 0.8  &  0.151  &  0.127 &  1.594    \\
 0.7  &  0.342  &  0.124 &  1.409    \\
 0.6  &  0.474  &  0.122 &  1.282   \\
 \hline
 \hline
$q$ & $\delta_{q}$ & $\eta_{q}$ & $\nu_{q}$   \\
 \hline
 1\cite{PhysRevD.103.116024}   &  18.200  &  0.208 &  1.333          \\
 0.9  &  15.327  &  0.245 &  1.072    \\
 0.8  &  13.545  &  0.275 &  0.924    \\
 0.7  &  12.333  &  0.300 &  0.829    \\
 0.6  &  11.500  &  0.320 &  0.763   \\
 \hline
 \hline
$q$ & $\sigma_{q}$ & $\tau_{q}$ & $\Omega_{q}$   \\
 \hline
 1\cite{PhysRevD.103.116024}   &  0.396  &  2.055 &  0.791          \\
 0.9  &  0.497  &  2.065 &  0.799    \\
 0.8  &  0.581  &  2.074 &  0.805    \\
 0.7  &  0.652  &  2.081 &  0.811    \\
 0.6  &  0.712  &  2.087 &  0.815   \\
 \hline 
\end{tabular}
\begin{tabular}{ p{1.4cm}p{1.2cm}p{1.2cm}p{1.2cm}p{1.2cm}p{1.5cm}  }
 \hline
 $q$ & $1$\cite{PhysRevD.103.116024} & $0.9$ & $0.8$ & $0.7$ & $0.6$   \\
 \hline

 $\omega_{q}$ &  1.500 &  1.500 & 1.500 &  1.500  &  1.500    \\
 \hline
 \hline
\end{tabular}
\label{exponentsP2}
\end{table}

\par In tables \ref{exponentsP3}-\ref{exponentsP5} we display the the numerical values for the nonextensive critical exponents for the percolation problem for $3d$, $4d$ and $5d$ systems. The referred extensive critical exponents values employed are that of Ref. \cite{PhysRevD.103.116024} from five-loop expansion. 

\begin{table}[H]
\caption{Nonextensive critical exponents, for some values of $q$, to $3$d nonextensive systems in percolation, obtained from NSFT (extensive indices values from cKP17-2 estimates \cite{PhysRevD.103.116024}).}
\begin{tabular}{ p{1.4cm}p{2.2cm}p{2.2cm}p{2.2cm}  }
 \hline
 \hline
 $q$ & $\alpha_{q}$ & $\beta_{q}$ & $\gamma_{q}$   \\
 \hline
 1\cite{PhysRevD.103.116024}   &  -0.64(6)  &  0.43(1) &  1.79(4)          \\
 0.9  &  -0.37(6)  &  0.41(1) &  1.58(2)    \\
 0.8  &  -0.16(6)  &  0.40(1) &  1.43(2)    \\
 0.7  &  -0.04(6)  &  0.38(1) &  1.33(2)    \\
 0.6  &   0.08(6)  &  0.37(1) &  1.25(2)   \\
 \hline
 \hline
$q$ & $\delta_{q}$ & $\eta_{q}$ & $\nu_{q}$   \\
 \hline
 1\cite{PhysRevD.103.116024}   &  5.19(6)  &  -0.03(1) & 0.88(2)          \\
 0.9  &  5.00(6)  &  0.00(1) &  0.79(1)    \\
 0.8  &  4.88(6)  &  0.02(1) &  0.72(1)    \\
 0.7  &  4.77(6)  &  0.04(1) &  0.68(1)    \\
 0.6  &  4.71(5)  &  0.05(1) &  0.64(1)   \\
 \hline
 \hline
$q$ & $\sigma_{q}$ & $\tau_{q}$ & $\Omega_{q}$   \\
 \hline
 1\cite{PhysRevD.103.116024}   &  0.45(1)  &  2.19(0) &  0.55(4)          \\
 0.9  &  0.51(1)  &  2.20(0) &  0.56(4)    \\
 0.8  &  0.56(1)  &  2.20(0) &  0.56(4)    \\
 0.7  &  0.59(1)  &  2.21(0) &  0.56(4)    \\
 0.6  &  0.63(1)  &  2.21(0) &  0.56(4)   \\
 \hline 
\end{tabular}
\begin{tabular}{ p{1.4cm}p{1.2cm}p{1.2cm}p{1.2cm}p{1.2cm}p{1.5cm}  }
 \hline
 $q$ & $1$\cite{PhysRevD.103.116024} & $0.9$ & $0.8$ & $0.7$ & $0.6$   \\
 \hline
 $\omega_{q}$ &  1.39(9) &  1.39(9) & 1.39(9) &  1.39(9)  &  1.39(9)    \\
 \hline
 \hline
\end{tabular}
\label{exponentsP3}
\end{table}

\begin{table}[H]
\caption{Nonextensive critical exponents, for some values of $q$, to $4$d nonextensive systems in percolation, obtained from NSFT (extensive indices values from cKP17-2 estimates \cite{PhysRevD.103.116024}).}
\begin{tabular}{ p{1.4cm}p{2.2cm}p{2.2cm}p{2.2cm}  }
 \hline
 \hline
 $q$ & $\alpha_{q}$ & $\beta_{q}$ & $\gamma_{q}$   \\
 \hline
 1\cite{PhysRevD.103.116024}   &  -0.744(8)  &  0.657(2) &  1.430(5)          \\
 0.9  &  -0.580(8)  &  0.639(2) &  1.333(3)    \\
 0.8  &  -0.464(8)  &  0.625(2) &  1.263(3)    \\
 0.7  &  -0.372(8)  &  0.614(2) &  1.201(3)    \\
 0.6  &  -0.300(8)  &  0.605(2) &  1.167(3)   \\
 \hline
 \hline
$q$ & $\delta_{q}$ & $\eta_{q}$ & $\nu_{q}$   \\
 \hline
 1\cite{PhysRevD.103.116024}   &  3.175(9)  &  -0.084(4) &  0.686(2)          \\
 0.9  &  3.137(9)  &  -0.066(4) &  0.645(1)    \\
 0.8  &  3.105(8)  &  -0.051(4) &  0.616(1)    \\
 0.7  &  3.077(8)  &  -0.038(4) &  0.593(1)    \\
 0.6  &  3.061(8)  &  -0.030(4) &  0.575(1)   \\
 \hline
 \hline
$q$ & $\sigma_{q}$ & $\tau_{q}$ & $\Omega_{q}$   \\
 \hline
 1\cite{PhysRevD.103.116024}   &  0.479(1)  &  2.315(1) &  0.38(3)          \\
 0.9  &  0.511(1)  &  2.319(1) &  0.38(3)    \\
 0.8  &  0.537(1)  &  2.322(1) &  0.38(3)    \\
 0.7  &  0.559(1)  &  2.325(1) &  0.38(3)    \\
 0.6  &  0.577(1)  &  2.327(1) &  0.38(3)   \\
 \hline 
\end{tabular}
\begin{tabular}{ p{1.4cm}p{1.2cm}p{1.2cm}p{1.2cm}p{1.2cm}p{1.5cm}  }
 \hline
 $q$ & $1$\cite{PhysRevD.103.116024} & $0.9$ & $0.8$ & $0.7$ & $0.6$   \\
 \hline 
 $\omega_{q}$ &  1.14(9) &  1.14(9) & 1.14(9) &  1.14(9)  &  1.14(9)    \\
 \hline
 \hline
\end{tabular}
\label{exponentsP4}
\end{table}
\begin{table}[H]
\caption{Nonextensive critical exponents, for some values of $q$, to $5$d nonextensive systems in percolation, obtained from NSFT (extensive indices values from cKP17-2 estimates \cite{PhysRevD.103.116024}).}
\begin{tabular}{ p{1.4cm}p{2.2cm}p{2.2cm}p{2.2cm}  }
 \hline
 \hline
 $q$ & $\alpha_{q}$ & $\beta_{q}$ & $\gamma_{q}$   \\
 \hline
 1\cite{PhysRevD.103.116024}   &  -0.865(0)  &  0.844(0) &  1.177(1)          \\
 0.9  &  -0.790(0)  &  0.838(0) &  1.141(1)    \\
 0.8  &  -0.735(0)  &  0.833(0) &  1.114(1)    \\
 0.7  &  -0.690(0)  &  0.830(0) &  1.093(1)    \\
 0.6  &  -0.650(0)  &  0.825(0) &  1.076(1)   \\
 \hline
 \hline
$q$ & $\delta_{q}$ & $\eta_{q}$ & $\nu_{q}$   \\
 \hline
 1\cite{PhysRevD.103.116024}   &  2.394(1)  &  -0.054(1) &  0.573(0)          \\
 0.9  &  2.384(1)  &  -0.045(1) &  0.558(0)    \\
 0.8  &  2.375(1)  &  -0.037(1) &  0.547(0)    \\
 0.7  &  2.368(1)  &  -0.031(1) &  0.538(0)    \\
 0.6  &  2.367(1)  &  -0.030(1)  & 0.530(0)   \\
 \hline
 \hline
$q$ & $\sigma_{q}$ & $\tau_{q}$ & $\Omega_{q}$   \\
 \hline
 1\cite{PhysRevD.103.116024}   &  0.495(0)  &  2.418(0) &  0.20(1)          \\
 0.9  &  0.509(0)  &  2.419(0) &  0.20(1)    \\
 0.8  &  0.520(0)  &  2.421(0) &  0.20(1)    \\
 0.7  &  0.529(0)  &  2.422(0) &  0.20(1)    \\
 0.6  &  0.537(0)  &  2.422(0) &  0.20(1)   \\
 \hline 
\end{tabular}
\begin{tabular}{ p{1.4cm}p{1.2cm}p{1.2cm}p{1.2cm}p{1.2cm}p{1.5cm}  }
 \hline
 $q$ & $1$\cite{PhysRevD.103.116024} & $0.9$ & $0.8$ & $0.7$ & $0.6$   \\
 \hline
 $\omega_{q}$ &  0.71(3) &  0.71(3) & 0.71(3) &  0.71(3)  &  0.71(3)    \\
 \hline
 \hline
\end{tabular}
\label{exponentsP5}
\end{table}

\subsection{Lee-Yang edge singularity}

\par For the Lee-Yang edge singularity problem, the critical indices $\eta_{q}$ and $\nu_{q}$ are not independent \cite{Bonfirm_1981,PhysRevD.95.085001}. In fact, they are related through $\nu_{q} = 2(d - 2 + \eta_{q})^{-1}$ \cite{Bonfirm_1981,PhysRevD.95.085001}. We compute $\eta_{q}$ with $\alpha = -1$ and $\beta = -1$ \cite{Bonfirm_1981} and then compute $\nu_{q}$ from the relation just mentioned for $\nu_{q}$ in terms of $\eta_{q}$. After evaluating these critical indices, we can compute the remaining ones through the scaling relations among them \cite{PhysRevD.103.116024}. In table \ref{exponentsLY2} we present the results for the numerical values for the nonextensive critical exponents for the Lee-Yang edge singularity problem for $2d$ systems. For that, we apply the corresponding extensive exponents values obtained from the exact solution \cite{PhysRevD.103.116024}. In this case, once again, these results for $2d$ systems are \emph{exact}. As there is no exact value of the extensive critical exponent $\omega$ for two dimensions for the Lee-Yang edge singularity problem, we have not to evaluate the corresponding nonextensive values $\omega_{q}$ as well as for $\Omega_{q}$ exponent which depends on $\omega_{q}$.

\begin{table}[H]
\caption{Exact nonextensive critical exponents, for some values of $q$, to $2$d nonextensive Lee-Yang edge singularity, obtained from NSFT. The exponent $\beta_{q}$ varies very slowly with $q$. It is constant in the approximation level shown.}
\begin{tabular}{ p{1.4cm}p{2.2cm}p{2.2cm}p{2.2cm}  }
 \hline
 \hline
 $q$ & $\alpha_{q}$ & $\beta_{q}$ & $\gamma_{q}$   \\
 \hline
 1\cite{PhysRevD.103.116024}   &  7.000  &  1.000 &  -7.000          \\
 0.9  &  7.650  &  1.000 &  -7.650    \\
 0.8  &  8.290  &  1.000 &  -8.290    \\
 0.7  &  8.920  &  1.000 &  -8.920    \\
 0.6  &  9.562  &  1.000 &  -9.562   \\
 \hline
 \hline
$q$ & $\delta_{q}$ & $\eta_{q}$ & $\nu_{q}$   \\
 \hline
 1\cite{PhysRevD.103.116024}   &  -6.000  &  -0.800 &  -2.500          \\
 0.9  &  -6.650  &  -0.708 &  -2.825    \\
 0.8  &  -7.289  &  -0.636 &  -3.145    \\
 0.7  &  -7.920  &  -0.578 &  -3.460    \\
 0.6  &  -8.561  &  -0.529 &  -3.781   \\
 \hline
 \hline
$q$ & $\sigma_{q}$ & $\tau_{q}$ &    \\
 \hline
 1\cite{PhysRevD.103.116024}   &  -0.167  &  1.833 &           \\
 0.9  &  -0.150  &  1.850 &      \\
 0.8  &  -0.137  &  1.863 &      \\
 0.7  &  -0.126  &  1.874 &      \\
 0.6  &  -0.117  &  1.883 &     \\
 \hline 
\end{tabular}
\label{exponentsLY2}
\end{table}

\par In tables \ref{exponentsLY3}-\ref{exponentsLY5} we show the the numerical values for the nonextensive critical exponents for the Lee-Yang edge singularity problem for $3d$, $4d$ and $5d$ systems. Once again, the referred extensive critical exponents values employed are that of Ref. \cite{PhysRevD.103.116024} from five-loop expansion. As argued in \cite{PhysRevD.103.116024}, the study of the extensive exponent $\omega$ for the Lee-Yang edge singularity problem has not produced any conclusions. So we have not to compute their corresponding nonextensive values $\omega_{q}$ for the Lee-Yang edge singularity problem as well as for the $\Omega_{q}$ exponent which depends on $\omega_{q}$.

\begin{table}[H]
\caption{Nonextensive critical exponents, for some values of $q$, to $3$d nonextensive Lee-Yang edge singularity, obtained from NSFT (extensive indices values from cKP17-2 estimates \cite{PhysRevD.103.116024}).}
\begin{tabular}{ p{1.4cm}p{2.2cm}p{2.2cm}p{2.2cm}  }
 \hline
 \hline
 $q$ & $\alpha_{q}$ & $\beta_{q}$ & $\gamma_{q}$   \\
 \hline
 1\cite{PhysRevD.103.116024}   &  -11.545(276)  &  1.000(29) &  11.545(239)          \\
 0.9  &  -9.703(207)  &  0.999(25) &  9.706(175)    \\
 0.8  &  -8.602(168)  &   1.000(23) &  8.602(140)    \\
 0.7  &  -7.837(144)  &   1.000(21) &  7.837(119)    \\
 0.6  &  -7.288(129)  &   1.000(20) &  7.288(105)   \\
 \hline
 \hline
$q$ & $\delta_{q}$ & $\eta_{q}$ & $\nu_{q}$   \\
 \hline
 1\cite{PhysRevD.103.116024}   &  12.544(275)  &  -0.557(9) &  4.515(92)          \\
 0.9  &  10.719(206)  &  -0.488(9) &  3.901(69)    \\
 0.8  &  9.601(169)  &  -0.434(9) &  3.534(56)    \\
 0.7  &  8.836(145)  &  -0.390(9) &  3.279(48)    \\
 0.6  &  8.288(129)  &   -0.354(9) &  3.096(43)   \\
 \hline
 \hline
$q$ & $\sigma_{q}$ & $\tau_{q}$ &    \\
 \hline
 1\cite{PhysRevD.103.116024}   &  0.080(1)  &  2.080(2) &            \\
 0.9  &  0.093(2)  &  2.093(2) &      \\
 0.8  &  0.104(2)  &  2.104(2) &      \\
 0.7  &  0.113(2)  &  2.113(2) &      \\
 0.6  &  0.121(2)  &  2.121(2) &     \\
 \hline 
\end{tabular}
\label{exponentsLY3}
\end{table}

\begin{table}[H]
\caption{Nonextensive critical exponents, for some values of $q$, to $4$d nonextensive Lee-Yang edge singularity, obtained from NSFT (extensive indices values from cKP17-2 estimates \cite{PhysRevD.103.116024}).}
\begin{tabular}{ p{1.4cm}p{2.2cm}p{2.2cm}p{2.2cm}  }
 \hline
 \hline
 $q$ & $\alpha_{q}$ & $\beta_{q}$ & $\gamma_{q}$   \\
 \hline
 1\cite{PhysRevD.103.116024}   &  -2.812(24)  &  1.000(7) &  2.813(17)          \\
 0.9  &  -2.684(20)  &  1.000(6) &  2.684(15)    \\
 0.8  &  -2.588(20)  &  1.000(6) &  2.588(15)    \\
 0.7  &  -2.512(20)  &  1.000(6) &  2.512(14)    \\
 0.6  &  -2.452(20)  &  1.000(6) &  2.452(14)   \\
 \hline
 \hline
$q$ & $\delta_{q}$ & $\eta_{q}$ & $\nu_{q}$   \\
 \hline
 1\cite{PhysRevD.103.116024}   &  3.813(23)  &  -0.338(8) &  1.203(6)          \\
 0.9  &  3.684(22)  &  -0.292(8) &  1.171(5)    \\
 0.8  &  3.587(21)  &  -0.256(8) &  1.147(5)    \\
 0.7  &  3.512(20)  &  -0.227(8) &  1.128(5)    \\
 0.6  &  3.452(20)  &  -0.203(8) &  1.113(5)   \\
 \hline
 \hline
$q$ & $\sigma_{q}$ & $\tau_{q}$ &   \\
 \hline
 1\cite{PhysRevD.103.116024}   &  0.262(1)  &  2.262(2) &            \\
 0.9  &  0.271(1)  &  2.271(2) &      \\
 0.8  &  0.279(1)  &  2.279(2) &      \\
 0.7  &  0.285(1)  &  2.285(2) &      \\
 0.6  &  0.290(1)  &  2.290(2) &     \\
 \hline 
\end{tabular}
\label{exponentsLY4}
\end{table}

\begin{table}[H]
\caption{Nonextensive critical exponents, for some values of $q$, to $5$d nonextensive Lee-Yang edge singularity, obtained from NSFT (extensive indices values from cKP17-2 estimates \cite{PhysRevD.103.116024}).}
\begin{tabular}{ p{1.4cm}p{2.2cm}p{2.2cm}p{2.2cm}  }
 \hline
 \hline
 $q$ & $\alpha_{q}$ & $\beta_{q}$ & $\gamma_{q}$   \\
 \hline
 1\cite{PhysRevD.103.116024}   &  -1.505(5)  &  1.000(2) &  1.506(3)          \\
 0.9  &  -1.480(5)  &  1.001(2) &  1.479(3)    \\
 0.8  &  -1.455(5)  &  1.000(2) &  1.456(3)    \\
 0.7  &  -1.440(5)  &  1.000(2) &  1.439(3)    \\
 0.6  &  -1.425(5)  &  1.000(2) &  1.425(3)   \\
 \hline
 \hline
$q$ & $\delta_{q}$ & $\eta_{q}$ & $\nu_{q}$   \\
 \hline
 1\cite{PhysRevD.103.116024}   &  2.506(4)  &  -0.148(3) &  0.701(1)          \\
 0.9  &  2.478(4)  &  -0.125(3) &  0.696(1)    \\
 0.8  &  2.47574)  &  -0.107(3) &  0.691(1)    \\
 0.7  &  2.439(4)  &  -0.092(3) &  0.688(1)    \\
 0.6  &  2.425(4)  &  -0.080(3) &  0.685(1)   \\
 \hline
 \hline
$q$ & $\sigma_{q}$ & $\tau_{q}$ &   \\
 \hline
 1\cite{PhysRevD.103.116024}   &  0.399(1)  &  2.399(1) &            \\
 0.9  &  0.403(1)  &  2.404(1) &      \\
 0.8  &  0.407(1)  &  2.407(1) &      \\
 0.7  &  0.410(1)  &  2.410(1) &      \\
 0.6  &  0.412(1)  &  2.412(1) &     \\
 \hline 
\end{tabular}\label{exponentsLY5}
\end{table}

\section{Conclusions}

\par We have employed tools from the recently introduced NSFT for computing the nonextensive critical exponents values for nonextensive $\lambda\phi^{3}$ scalar field theories for all loop orders and $|q - 1| < 1$. These values are composed of three-level, two-loop order and higher ones which are extensive (which result from considering all length scales) and a one-loop level $q$-dependent term (originating from small length scales), where the latter is a slight one. We have applied the results for the specific problems of percolation and Lee-Yang edge singularity valid in the nonextensive realm. The 2d results are \emph{exact}, within the approximation employed in this work. The results for $3$d, $4$d and $5$d are also obtained, where the corresponding extensive values are that obtained from five-loop resummation. The nonextensive critical indices depend on the nonextensive parameter $q$  and reduce to their corresponding extensive values in the limit $q\rightarrow 1$. The results show an interplay between global correlations and fluctuations, since the nonextensive critical indices values depend on $q$. Such dependence is in agreement with the universality hypothesis, once these global correlations are interpreted as effective interactions, so the corresponding critical exponents must depend on $q$. The present work opens a new road for many further works using both computational and experimental methods for trying to agreeing with their results with that shown in this work. 

%

\section*{Acknowledgments} 

\par PRSC would like to thank the Brazilian funding agencies CAPES and CNPq (grants: Universal-431727/2018 and Produtividade 307982/2019-0) for financial support. 

\bibliography{apstemplate}

\end{document}